\documentclass[10pt, conference]{IEEEtran} 

\usepackage{times}
\usepackage{tikz}
\usetikzlibrary{positioning}
\usepackage{pgfplotstable}
\pgfplotsset{compat=1.3}
\usepgfplotslibrary{fillbetween}
\usetikzlibrary{patterns}
\usepackage{subfiles}
\usepackage{xcolor}
\usepackage{cite}
\usepackage{amsmath,amssymb,amsfonts}
\usepackage{algorithmic}
\usepackage{graphicx}
\usepackage{textcomp}
\usepackage{xcolor}
\usepackage{booktabs}
\usepackage{multirow}
\pgfmathdeclarefunction{gauss}{2}{%
  \pgfmathparse{1/(#2*sqrt(2*pi))*exp(-((x-#1)^2)/(2*#2^2))*0.7}%
}

\def\BibTeX{{\rm B\kern-.05em{\sc i\kern-.025em b}\kern-.08em
    T\kern-.1667em\lower.7ex\hbox{E}\kern-.125emX}}
\pgfplotsset{compat=1.15}

\usepackage{subfigure}
\usepackage{epstopdf}
\usepackage{epsfig}
\usepackage{array}
\usepackage{lipsum}
\usepackage[]{algorithm2e}
\usepackage{tabularx}
\usepackage{hyperref}

\usepackage{enumitem}

\usepackage{booktabs} 

\usepackage{graphicx}
\usepackage{booktabs}
\usepackage{multirow}
\usepackage{multicol}
\usepackage{array}
\usepackage{tabularx}
\usepackage{bm}
\usepackage{soul}
\usepackage{tikz}
\usepackage{amssymb}
\usepackage{amsmath}
\usepackage{xfrac}
\usepackage[bottom]{footmisc}

\def\invcircledast#1{%
  \mathbin{\vphantom{\circledast}\text{%
    \ooalign{\smash{\blackcircle}\cr
             \hidewidth\smash{\textcolor{white}{\bf \footnotesize $#1$}}\hidewidth\cr
            }%
  }}%
}

\newcommand{\blackcircle}{\raisebox{-.6ex}{\scalebox{2.30}{$\bullet$}}}

\usepackage{etoolbox}
\usepackage{bm}
\let\oldbibliography\thebibliography
\renewcommand{\thebibliography}[1]{\oldbibliography{#1}
\setlength{\itemsep}{-2.5pt}} 
\newcommand{\Design}{MIMHD\xspace}

\IEEEoverridecommandlockouts
\IEEEpubid{\makebox[\columnwidth]{978-1-6654-3922-0/21/\$31.00 $\copyright$2021 IEEE \hfill} \hspace{\columnsep}\makebox[\columnwidth]{ }}

\begin{document}


\title{\Design: Accurate and Efficient Hyperdimensional Inference Using Multi-Bit In-Memory Computing \vspace{-3mm}}

\author{Arman Kazemi$^{*\star}$, Mohammad Mehdi Sharifi$^{*}$, Zhuowen Zou$^{\psi}$, Michael Niemier$^{*}$, X. Sharon Hu$^{*}$, Mohsen Imani$^\dagger$\\
\normalsize $^{*}$University of Notre Dame, $^{\psi}$University of California San Diego,  \normalsize $^\dagger$University of California Irvine, $^{\star}$akazemi@nd.edu}

\maketitle

\begin{abstract}
Hyperdimensional Computing (HDC) is an emerging computational framework that mimics important brain functions by operating over high-dimensional vectors, called hypervectors (HVs). In-memory computing implementations of HDC are desirable since they can significantly reduce data transfer overheads. All existing in-memory HDC platforms consider binary HVs where each dimension is represented with a single bit. However, utilizing multi-bit HVs allows HDC to achieve acceptable accuracies in lower dimensions which in turn leads to higher energy efficiencies. Thus, we propose a highly accurate and efficient multi-bit in-memory HDC inference platform called \Design. \Design supports multi-bit operations using ferroelectric field-effect transistor (FeFET) crossbar arrays for multiply-and-add and FeFET multi-bit content-addressable memories for associative search. We also introduce a novel hardware-aware retraining framework (HWART) that trains the HDC model to learn to work with \Design. For six popular datasets and 4000 dimension HVs, \Design using 3-bit (2-bit) precision HVs achieves (i) average accuracies of 92.6\% (88.9\%) which is 8.5\% (4.8\%) higher than binary implementations; (ii) 84.1$\times$ (78.6$\times$) energy improvement over a GPU, and (iii) 38.4$\times$ (34.3$\times$) speedup over a GPU, respectively. The 3-bit \Design is 4.3$\times$ and 13$\times$ faster and more energy-efficient than binary HDC accelerators while achieving similar accuracies.
\end{abstract}


\IEEEpeerreviewmaketitle

\section{Introduction}

Hyperdimensional Computing (HDC) is an emerging computational framework based on the observation that the brain operates in hyperdimensional spaces over high-dimensional vectors to perform cognition and perception~\cite{kanerva2009hyperdimensional}. The computational units of HDC are high-dimensional vectors called hypervectors (HVs). HDC is applicable to a wide range of applications such as robotics~\cite{mitrokhin2019learning}, machine learning, and cognitive computing. HDC achieves similar or higher accuracy than convolutional neural networks (CNNs), support vector machines (SVMs), and extreme gradient boosting for applications such as distributed sensors~\cite{kleyko2018classification}, multimodal sensor fusion\cite{Rasanen15}, and biomedical signal processing~\cite{rahimi2018efficient}.

Another important advantage of HDC over other machine learning models such as CNNs and SVMs is the higher energy efficiency\cite{imani2019framework, imani2021revisiting}. HDC operates over high-dimensional HVs and can greatly benefit from the high parallelism offered by processing in-memory (PIM) architectures and offer even more energy and latency improvements~\cite{li2016hyperdimensional, imani2017exploring}. In addition, the inherent robustness of HDC to computational errors makes it amenable to PIM architectures based on emerging devices with variability issues. However, previous PIM HDC works in the literature has focused on binary implementations of HDC which suffer from accuracy loss compared to floating point models and require dimensions higher that 10,000 which thwarts achieving high energy efficiencies~\cite{imani2019quanthd, cano2021onlinehd}.

We observe that using 2 or 3-bit HVs could improve HDC's inference accuracy to match that of a high precision model. This is inline with the fact that the synapses in the brain have 4.7 bits of precision~\cite{bartol2015nanoconnectomic}. Furthermore, using multi-bit HVs allows reducing the number of dimensions at iso-accuracy which in turn offers significant energy savings. One of the most prominent impediments to designing multi-bit HDC accelerators is associative search based on cosine distance. Binary HDC implementations leverage Hamming distance measurements using ternary content-addressable memories (TCAMs) to avoid cosine distance. On the other hand, multiple binary storage elements are required to store multi-bit HVs which reduces energy efficiency gains and adds more complexity.

Recently, ferroelectric field-effect transistor(FeFET) multi-bit content-addressable memories (MCAMs) were proposed that can perform single-step associative search based on the MCAM distance metric~\cite{kazemi2021memory}. Furthermore, multi-level FeFETs for storage and computation have been experimentally demonstrated~\cite{jerry2018ferroelectric}. These advances make FeFETs a suitable candidate for designing PIM based multi-bit HDC accelerators.

In this paper, we introduce \Design, an in-memory implementation of HDC inference. \Design achieves comparable accuracy as high-bit precision HDC while offering significant energy and latency reduction. \Design directly represents each dimension of a HV using multi-bit FeFETs and supports essential HDC inference operations, i.e., encoding and associative search. \Design uses FeFET crossbar arrays~\cite{jerry2018ferroelectric} to implement encoding, and FeFET MCAMs~\cite{kazemi2021memory} for associative search. We propose a novel hardware-aware retraining (HWART) algorithm that trains the HDC model to compensate for possible accuracy loss due to the MCAM distance metric. We evaluate the efficiency and accuracy of \Design over a wide range of classification problems. For six popular datasets, \Design using 2 and 3-bit precisions achieves (i) average accuracies of 88.9\% and 92.6\% which is similar to high precision (8-bit) implementations, (ii) 84.1$\times$ and 78.6$\times$ energy improvement over a GPU, and (iii) 38.4$\times$ and 34.3$\times$ speedup over a GPU, respectively. 3-bit \Design is 4.3$\times$ and 13$\times$ faster and more energy-efficient than the binary HDC accelerators~\cite{imani2019searchd}.

\section{Background and Related Work}
\subsection{Hyperdimensional Computing Basics}
HDC is a neuro-inspired model of computation based on the observation that the human brain operates on high-dimensional and distributed representations of data~\cite{kanerva2009hyperdimensional} called hypervectors (HVs). Researchers have applied HDC to diverse cognitive tasks, such as robotics~\cite{mitrokhin2019learning}, classification~\cite{nazemi2020synergiclearning}, bio-signal processing~\cite{Poduval2021cognitive}, and regression~\cite{Cano2021reghd}. Most of the previous work use binary HVs. Below, we discuss an overview of HDC.  

\begin{figure}[t!]
\centering
\epsfig{file=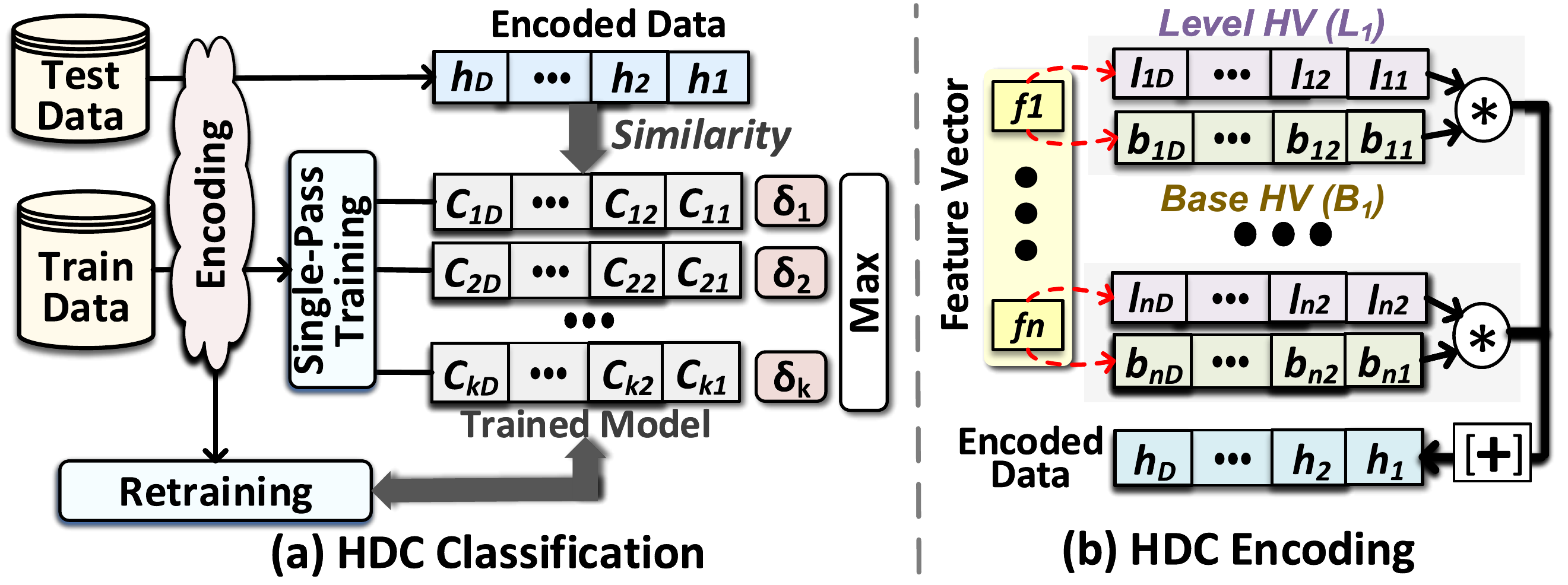, width=1\columnwidth}
\vspace{-6mm}
\caption{(a) An overview of HDC classification, (b) HDC encoding module.}
\vspace{-4mm}
\label{fig:HD}
\end{figure} 

\subsubsection{\textbf{HDC Encoding}}\label{sec:encoding_bg}
The first step in HDC is to map each data point into high-dimensional space. The mapping procedure is often referred to as \textit{encoding} (shown in Fig.~\ref{fig:HD}b). The encoding is such that the data points different from each other in the original space are also different in the HDC space. 
To explain the encoding process, assume an input feature vector (an image, voice, etc.) in the original space $\vec{{F}} = \{f_1,~f_2,\cdots,f_n\}$ and $\vec{F} \in \mathcal{R}^n$. HDC first linearly quantizes the feature vector values into $m$ levels, $\{\bar{q}_1, \bar{q}_2, \cdots, \bar{q}_m$\}, and assigns a level HV to each quantization level, $\{\vec{L}_1, \vec{L}_2, \cdots, \vec{L}_m\}$. The level HVs are generated such that $\vec{L}_1$ and $\vec{L}_m$ are orthogonal $\delta(\vec{{L}}_1, \vec{{L}}_m) \simeq 0$, but the intermediate levels follow a spectrum of similarity, depending on their physical closeness to $\bar{q}_1$ and $\bar{q}_m$. The encoding module maps $\vec{{F}}$ into a high-dimensional vector $\vec{\mathcal{H}} \in \{0,1\}^D$, where $\mathcal{D}>>n$. The following equation and Fig.~\ref{fig:HD}b show an encoding method that maps an input vector into high-dimensional space~\cite{imani2019framework}:

\begin{equation}\label{eq:encoding}
\vec{\mathcal{H}} = \sum_{k=1}^{n} {L(f_k)}* \vec{\mathcal{B}}_{k}
\end{equation}

\noindent where base HVs ($\vec{\mathcal{B}}_{k}$s) retain the spatial or temporal location of the features. Base HVs are randomly chosen, hence orthogonal. Base HVs are also of dimension $\mathcal{D}$. Function $L(f_k)$ assigns the corresponding level HV to the features.

\subsubsection{\textbf{HDC Training}}

To find the universal property for each class in the training dataset, the trainer module linearly combines HVs belonging to each class, i.e., adding the HVs to create a single HV for each class. Once all HVs are combined, we treat per-class accumulated HVs, called \textit{class HVs}, as the learned model. Fig.~\ref{fig:HD}a shows HDC operation during single-pass training. Assuming a problem with $k$ classes, $\mathcal{M}=\{\vec{\mathcal{C}_1}, \vec{\mathcal{C}_2}, \cdots, \vec{\mathcal{C}_k}\}$ represents the model. For example, after encoding all of training data belonging to class/label $l$, the class HV $\vec{\mathcal{C}^l}$ can be obtained by bundling (adding) all $\vec{\mathcal{H}}^l$s. Assuming there are $\mathcal{J}$ inputs with label $l$: $\vec{\mathcal{C}^l} = \sum_{j}^{\mathcal{J}}{\vec{\mathcal{H}^l_{j}}}$.

\subsubsection{\textbf{HDC Inference}}
For inference, the model checks the similarity of each encoded test data with the class HVs in two steps. The first step encodes the input (the same encoding used for training) to produce a query HV $\vec{\mathcal{H}}$. Then, as Fig.~\ref{fig:HD}a shows, we compute the similarity ($\delta$) of $\vec{\mathcal{H}}$ and all class HVs. Query data is labeled with the label of the class HV with the highest similarity.

\subsection{HDC Acceleration}

Prior work showed how the binary HDC models could be accelerated using PIM platforms~\cite{imani2019searchd, karunaratne2020memory}. For example, work in~\cite{imani2019searchd, li2016hyperdimensional, imani2017exploring} showed that PIM architectures leveraging CAMs can accelerate HDC associative search by two orders of magnitude. CAMs enable row-parallel associative search over binary HVs with the Hamming distance metric. However, using non-binary HVs cannot readily exploit current PIM architectures to accelerate HDC inference. A multi-bit PIM structure that supports associative search could be highly desirable for accelerating accurate HDC inference.

\subsection{Computing with Multi-Bit FeFETs}

\begin{figure}[t!]
\centering
\epsfig{file=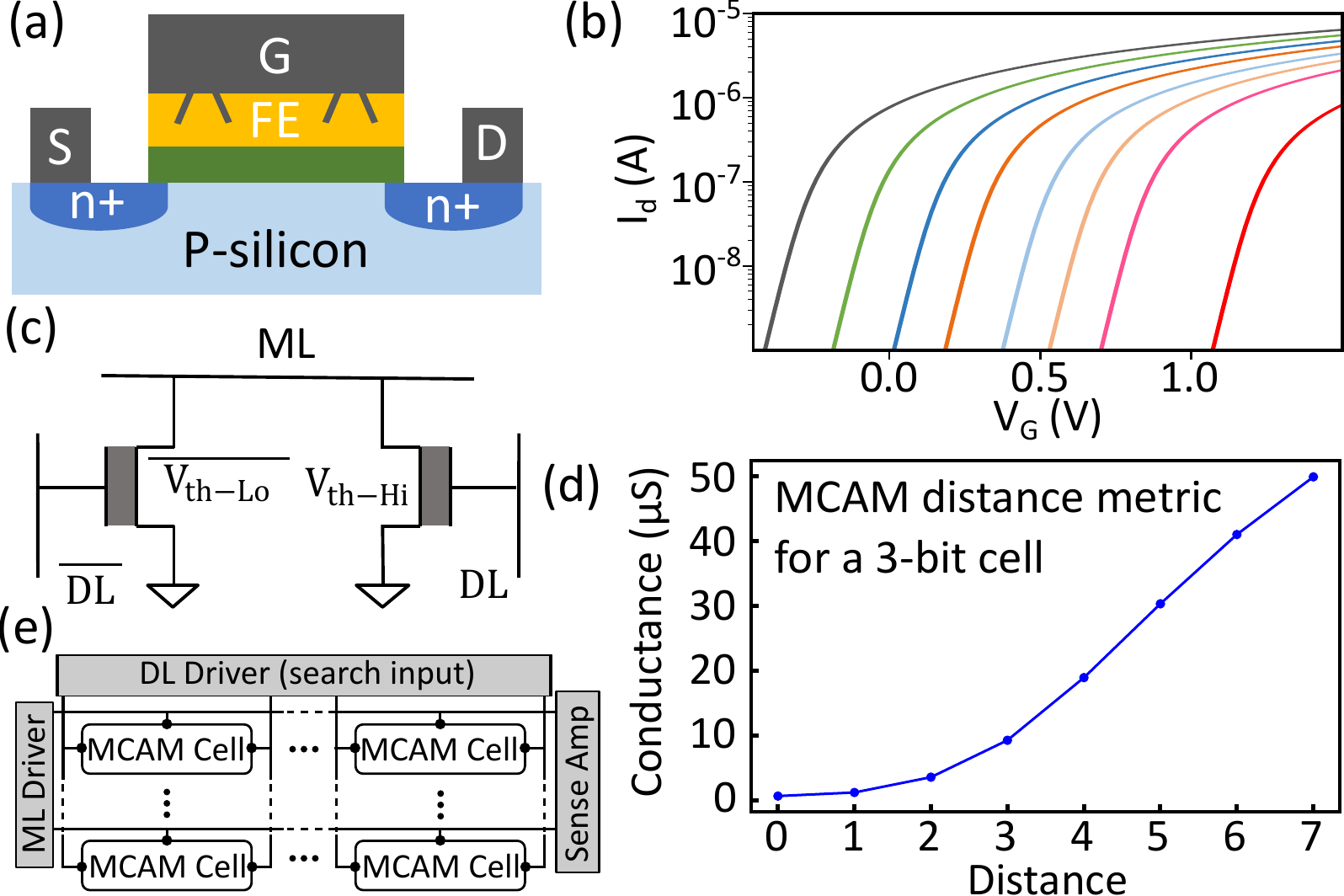, width=0.9\columnwidth}
\caption{(a) FeFET device schematic~\cite{ni2018circuit}; (b) Transfer characteristics of a FeFET programmed to 8 distinct states; (c) Schematic of the FeFET MCAM cell~\cite{kazemi2021memory}; (d) MCAM distance metric from~\cite{kazemi2021memory}; (e) MCAM array schematic.}
\vspace{-4mm}
\label{fig:fefet}
\end{figure}

The FeFET device is a promising candidate for PIM architectures. FeFETs are made by integrating a ferroelectric (FE) layer in the gate stack of a MOSFET~\cite{ni2018circuit} (Fig.~\ref{fig:fefet}a). The threshold voltage ($V_{th}$) of the FeFET is determined by the non-volatile polarization of the FE layer. Applying voltage pulses to the gate of the FeFET changes the polarization of the FE layer and consequently the $V_{th}$ of the FeFET device~\cite{jerry2018ferroelectric}. Fig.~\ref{fig:fefet}b shows the transfer characteristics of a FeFET device programmed to 8 different $V_{th}$ states, i.e., storing 3-bits of information~\cite{kazemi2021memory}. 

FeFET crossbar arrays have been used to accelerate neural network inference and training~\cite{kazemi2020hybrid,jerry2018ferroelectric}. Work in~\cite{kazemi2021memory} proposed a FeFET multi-bit CAM (MCAM), shown in Fig.~\ref{fig:fefet}c, that can perform parallel in-memory associative search based on the MCAM distance metric. FeFET MCAM cells store different states by programming $V_{th-Hi}$ and $\overline{V_{th-Lo}}$ and search using voltage signals on the data lines ($DL$ and $\overline{DL}$). Fig.~\ref{fig:fefet}d shows the MCAM distance metric for a single cell where the x-axis represents the distance between the stored state and the search input, and the y-axis indicates the conductance of the cell (which directly represents similarity). The conductance grows exponentially with respect to the distance and saturates towards the higher distance. This behavior results from the transfer characteristics of the FeFETs. By sensing the match-lines (ML in Fig.~\ref{fig:fefet}c), which are connected to all cells in a row (Fig.~\ref{fig:fefet}e), we can find the MCAM row with the lowest conductance (data most similar to the search query). Although the MCAM distance metric is not a previously known and commonly used metric like cosine or Euclidean, it achieves high accuracies for different applications~\cite{kazemi2021memory}. Furthermore, technologies like resistive RAM are amenable to multi-bit crossbar arrays, but they do not implement a CAM-based multi-bit associative search with a useful distance metric.

\section{Proposed \Design}

In this section, we present \Design, an accurate and efficient in-memory platform for HDC inference. We exploit multi-bit FeFET devices to directly support 1, 2, and 3-bit precision implementation of HDC. We leverage PIM to implement encoding and associative search operations to achieve high energy and latency efficiency. Furthermore, we introduce a novel hardware-aware retraining framework (HWART) that trains the HDC model to adapt to the unique features of FeFET MCAMs in \Design. The overall structure of \Design is shown in Fig.~\ref{fig:architecture} which consists of an encoding and an associative search module.

\subsection{Encoding}
As discussed in Section~\ref{sec:encoding_bg}, the goal of encoding is to transform the data into a high-dimensional space. Using a multi-bit encoding, we increase the information representation capacity of the HVs, compared to binary implementations. The encoding module of \Design operates in multi-bit precision and works with multi-bit level HVs and base HVs. To generate $m$ level HVs we generate a random $\mathcal{D}$-dimensional $P$-bit ($P \in \{1, 2, 3\}$ denotes the bit precision) HV to represent level 1. To generate the HV for level 2, we randomly choose $\mathcal{D}\slash m$ dimensions and randomly assign them values from $1$ to $2^P$. We continue this process until we have generated $m$ level HVs. Generating level HVs only happens once during training when $P$ and $m$ are chosen. As such, $f_k$'s in equation~\ref{eq:encoding} are represented with their corresponding level HVs. Base HVs are chosen randomly (Section~\ref{sec:encoding_bg}) and are also fixed.

\begin{figure}[t!]
\centering
\epsfig{file=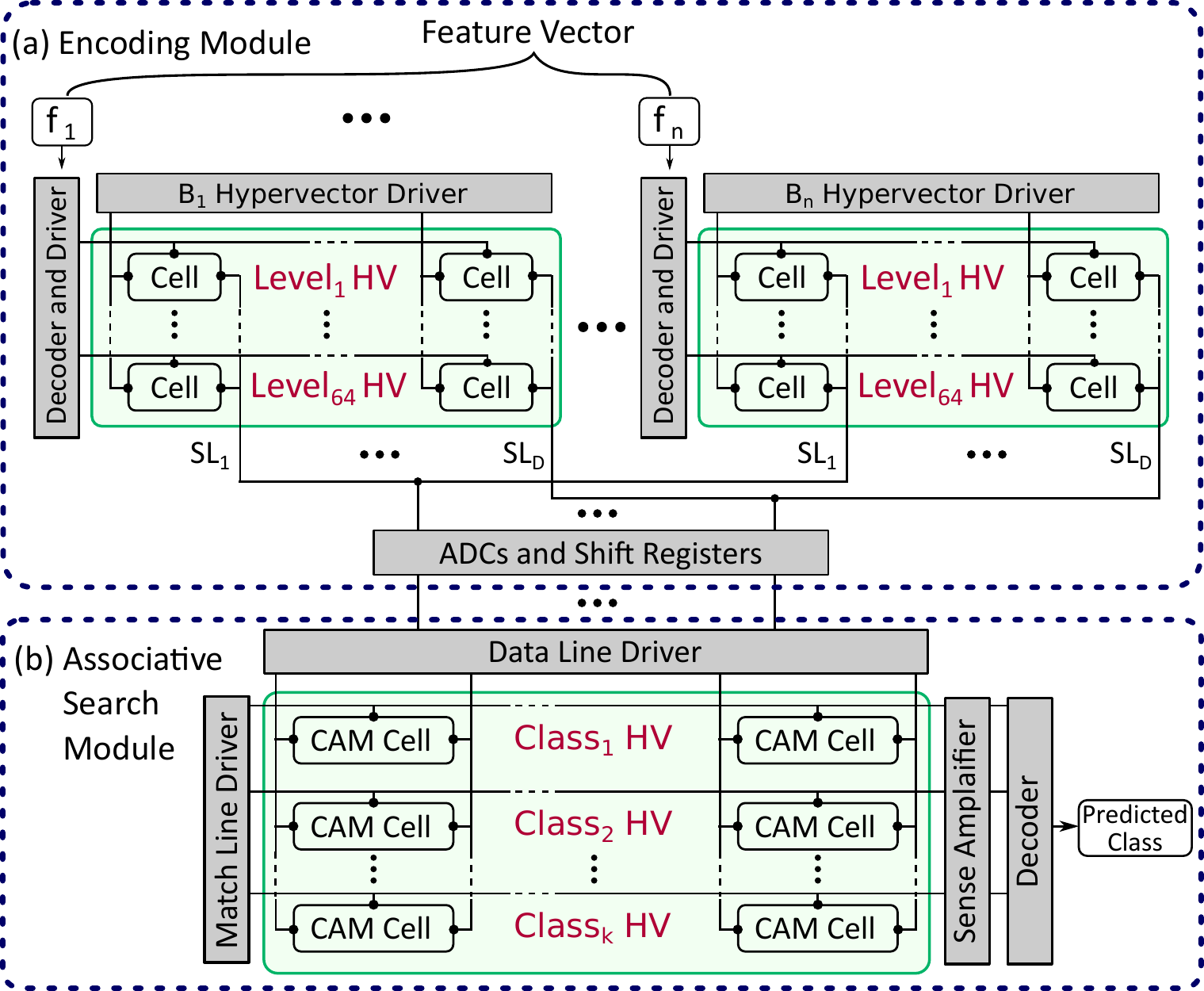, width=0.9\columnwidth}
\caption{\Design supports all required operations for HDC inference: (a) the encoding module consists of $n$ crossbar array groups with $\mathcal{D}/64$ crossbar arrays of size $64\times64$; (b) the associative search module consists of MCAM blocks that store $k$ class HVs.}
\vspace{-4mm}
\label{fig:architecture}
\end{figure}

Fig.~\ref{fig:architecture}a shows the encoding module of \Design. We utilize FeFET crossbar arrays to store the level HVs. We need to store a total of $n$ HVs of size $m \times\mathcal{D}$ for a dataset with $n$ features. We consider $m=64$ to support up to 64 level HVs and minimize accuracy loss due to HV quantization. Given that $\mathcal{D}$ is in the order of thousands, we use $\mathcal{D}/64$ crossbar arrays of size $64\times64$ for each feature since the computation is independent for each dimension. Each $i-th$ group of crossbar arrays ($i \in [1,2,\dots,n]$) activates the rows containing the level HV corresponding to $f_i$ and performs a multiplication between that level HV and the $i_{th}$ base HV ($\vec{\mathcal{B}}_{i}$) as its input. Source lines ($SL$s) add the resulting currents from these multiplications based on Kirchhoff's law and feed them to the analog-to-digital converters (ADCs). This current-based addition ensures that we do not need to store intermediate values during the encoding; avoiding expensive FeFET writes. The ADCs are used to generate a $P$-bit $\mathcal{D}$-dimensional encoded HV.

\subsection{Associative Search}
Associative search is the final step in HDC inference where the class HVs are compared with the query HV. Associative search is arguably the most challenging operation of multi-bit HDC to implement with PIM. TCAMs can realize associative search with Hamming distance, but they are only suitable for binary implementations. Thus, we utilize FeFET MCAMs~\cite{kazemi2021memory} for associative search in \Design. Given the high dimensionality of the class HVs, we use multiple MCAM arrays of size $64\times64$. The MCAMs store the $k$ trained class HVs in their rows (Fig.~\ref{fig:architecture}b). The encoded query is routed to the MCAMs as input to perform associative search with the stored class HVs based on the MCAM distance metric. We add the currents from the same rows of the different MCAMs to get the distance for $\mathcal{D}$ dimensions following the implementation in~\cite{imani2017exploring}. Then sense amplifies in~\cite{imani2019searchd} and used to detect the row that is the most similar to the encoded query (lowest current) and report the corresponding class as the result of the prediction.

The utilized MCAM cell (Fig.~\ref{fig:fefet}c) also supports Hamming distance search~\cite{ni2019ferroelectric}. This allows \Design to support the binary implementation of HDC by programming the FeFET crossbar arrays in a binary fashion. All peripherals that support multi-bit operations in \Design support the binary operations as well. The only difference between the binary and multi-bit operations is using Hamming distance for associative search for the former and the MCAM distance metric for the latter.

\begin{figure}[t!]
\centering
\epsfig{file=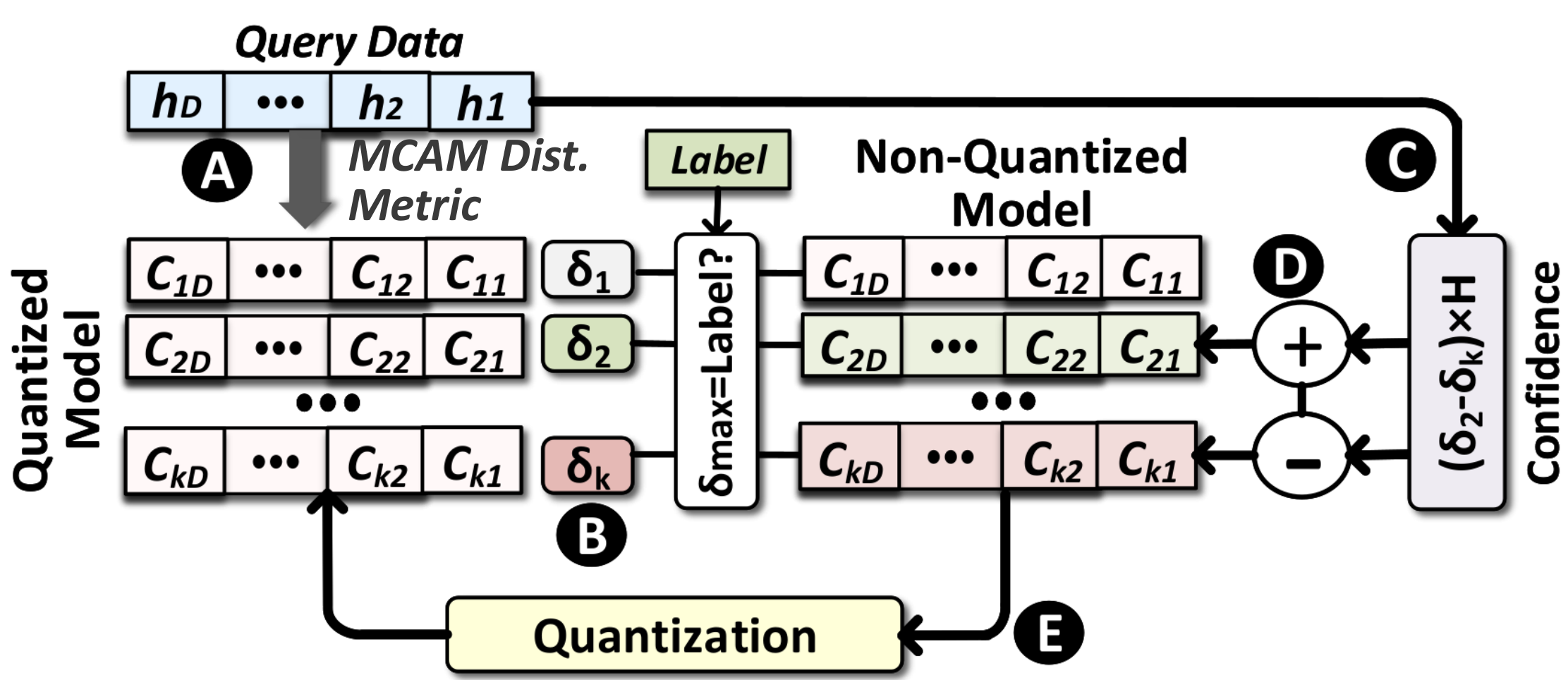, width=1\columnwidth}
\vspace{-4mm}
\caption{HWART framework adapts HDC models to work with the MCAM distance metric supported by \Design. }
\vspace{-3mm}
\label{fig:retraining}
\end{figure}

\subsection{Hardware-Aware Retraining (HWART)} \label{sec:retraining}

The HDC training method has been theoretically designed to work with the cosine distance. \Design uses the MCAM distance metric for associative search, which can result in a loss of classification accuracy. In order to compensate for this possible quality loss, we propose HWART, a retraining method that teaches the model to work with a new distance metric. 

Fig.~\ref{fig:retraining} shows an overview of the proposed HWART. We store two copies of the HDC model: quantized and non-quantized. The non-quantized model allows for a fine-grained accumulation of information and the quantized model mimics the forward pass of inference. During the following two phases, we teach the model to work with the quantization constraints: (i) feed-forward phase that checks the similarity of an encoded training data point with the quantized model using the MCAM distance metric, and (ii) model update phase that adjusts the non-quantized HDC model for each mispredicted sample. 
For each encoded training data point $\vec{\mathcal{H}}$, we check the similarity of $\vec{\mathcal{H}}$ with all class HVs stored in the quantized model to perform the prediction ($\invcircledast{A}$). This similarity search is based on the MCAM distance metric derived from the FeFET MCAM($\invcircledast{B}$). Then, we examine if the model returns the correct label $l$ for $\vec{\mathcal{H}}$. If the model mispredicts it as label $l'$, the non-quantized model is updated as follows ($\invcircledast{C}\invcircledast{D}$).

\begin{equation}\label{eq:retraining}
\begin{split}
\vec{\mathcal{C}_l} \gets \vec{\mathcal{C}_l} + \eta~ (\delta_{l'}-\delta_{l}) \times \mathcal{\vec{H}} \\
\vec{\mathcal{C}_{l'}} \gets \vec{\mathcal{C}_{l'}} - \eta~ (\delta_{l'}-\delta_{l}) \times \mathcal{\vec{H}}
\end{split}
\end{equation}

\noindent where $\delta_l = \delta(\vec{\mathcal{H}},\vec{\mathcal{C}_{l}})$ and $\delta_{l'} = \delta(\vec{\mathcal{H}},\vec{\mathcal{C}_{l'}})$ are the similarity of the query with correct and miss-predicted classes, respectively. Thus, we ensure that we update the model based on how ``far'' a training data point is miss-classified with the current model. In case of a very far miss-prediction, $\delta_{l'}>>\delta_{l}$, we make greater changes to the non-quantized model; while in case of a marginal miss-prediction, $\delta_{l'} \simeq \delta_{l}$, the update makes lesser changes to the non-quantized model.

\begin{figure}[t!]
\centering
\epsfig{file=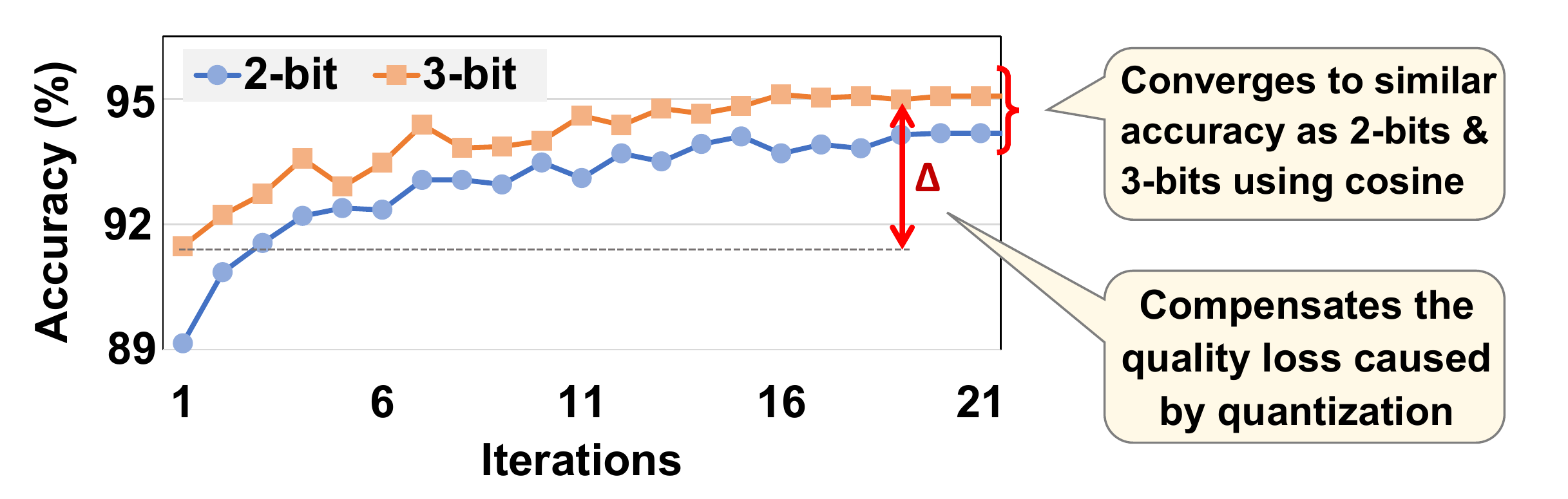, width=0.9\columnwidth}
\vspace{-3mm}
\caption{\Design accuracy during different HWART iterations.}
\vspace{-4mm}
\label{fig:HWAR_accuracy}
\end{figure}

After updating the non-quantized model over a batch of training data, we update the quantized model by mapping each model element of non-quantized model to the desired number of bits ($\invcircledast{E}$). We continue this adaptive iterative procedure until the classification accuracy stabilizes (Fig.~\ref{fig:HWAR_accuracy}).

\section{Evaluation}
\label{sec:eval}
\subsection{Experimental Setup}
We implement \Design both in software and hardware simulation. In software, we implemented and verified \Design functionality with a Python implementation. To evaluate \Design accuracy, we design a cycle-accurate simulator based on our framework in PyTorch~\cite{abadi2016tensorflow}. Our framework translates all PyTorch code directly to vector-based operations supported by our PIM architecture. For the hardware design, we use NeuroSim~\cite{peng2019dnn+} to obtain energy and latency results for the crossbar arrays including all required peripherals. We use HSPICE and the FeFETs Preisach model~\cite{ni2018circuit} to measure the energy consumption and latency of the FeFET MCAMs and other peripherals. All the hardware components are based on a 22nm technology node.

We compare the energy and latency consumption of \Design with a NVIDIA GTX 1080 GPU running the HDC models with 8-bit precision. 
The performance and energy of the GPU are measured by the \verb|nvidia-smi| tool. We evaluate the accuracy and efficiency of \Design on six popular datasets (listed in Table~\ref{tab:benchmark}), ranging from small datasets collected in a small IoT network to large datasets which includes hundreds of thousands of images of facial data. 
\begin{table}[t!]
\centering
\caption{Datasets ($n$: feature size, $k$: number of classes)}
\vspace{-2mm}
\label{tab:benchmark}
\resizebox{0.85\columnwidth}{!}{
\begin{tabular}{c|cccccc}
\toprule
& $n$ & $k$  & \shortstack{\textbf{Train}\\ \textbf{Size}} & \shortstack{\textbf{Test}\\ \textbf{Size}} & \textbf{Description}             \\ \midrule
\textbf{MNIST}  & 784               & 10                & 60,000              & 10,000             & Handwritten Recognition\cite{lecun1998gradient,ciregan2012multi}         \\ 
\textbf{UCIHAR} & 561               & 12                & 6,213                & 1,554               & Activity Recognition(Mobile)\cite{anguita2012human} \\ 
\textbf{ISOLET} & 617               & 26                & 6,238               & 1,559              & Voice Recognition~\cite{Dua:2019}              \\ 
\textbf{PAMAP} & 75                & 5              & 611,142             & 101,582            & Activity Recognition(IMU)~\cite{reiss2012introducing}    \\ 
\textbf{FACE}   & 608               & 2                & 522,441             & 2,494              & Face Recognition\cite{angelova2005pruning}\\  
\textbf{PECAN}   & 312  & 3     & 22,290    & 5,574 & Urban Electricity Prediction~\cite{shome}\\ \bottomrule
\end{tabular}
}
\vspace{-4mm}
\end{table}

\subsection{\Design Accuracy}
\Design accuracy depends on three parameters: (i) the HV dimensionality that determines the HV capacity and the level of redundancy, (ii) the precision of each HV element, and (iii) the distance metric used for associative search. 

\subsubsection{\textbf{Dimensionality}}
Fig.~\ref{fig:dimension}a compares HDC classification accuracy for different dimensionalities for face detection applications. Other datasets exhibit similar trends. Our evaluation shows that increasing dimensionality results in improving the classification accuracy for all configurations. In the low dimensional space, the binary model is much less accurate than the higher precision models. This accuracy gap shrinks by scaling the dimensionality. For example, the 1-bit precision model provides 17.9\% and 4.2\% lower accuracy than the 8-bit model in $\mathcal{D}{=}1k$ and $\mathcal{D}{=}10k$, respectively. Fig.~\ref{fig:dimension}b reports the maximum classification accuracy achieved by the HDC models using different bit precisions and the lowest dimensionality that provides such accuracy. The results are averaged over all tested datasets. 
Our evaluation shows that \Design, using multi-bit elements, reduces the number of required dimensions to achieve maximum accuracy compared to the binary model. \Design provides maximum classification accuracy in $\mathcal{D}{=}4k$ ($\mathcal{D}{=}8k$) for 3-bit (2-bit) precision. The results also indicate that, regardless of dimensionality, the binary HDC model using Hamming distance cannot achieve the maximum accuracy provided by the full precision (8-bit) model. In contrast, \Design can provide the same accuracy as the full precision model while enabling a fully in-memory implementation. 

\subsubsection{\textbf{Hypervector Precision}}
Fig.~\ref{fig:accuracy}a compares the classification accuracy of HDC models in different precisions using different search metrics/hardware. All results are reported using the same dimensionality ($\mathcal{D}{=}4k$). \Design is trained with HWART. Results show that HDC classification accuracy highly depends on model precision. \Design in 2-bit and 3-bit precisions provides, on average across all datasets, 8.3\% and 4.7\% higher classification accuracy as compared to the binary model. Moreover, 3-bit \Design can achieve the same accuracies as an 8-bit model using cosine distance. 

\subsubsection{\textbf{Distance Metric}}
\begin{figure}[t!]
\centering
\epsfig{file=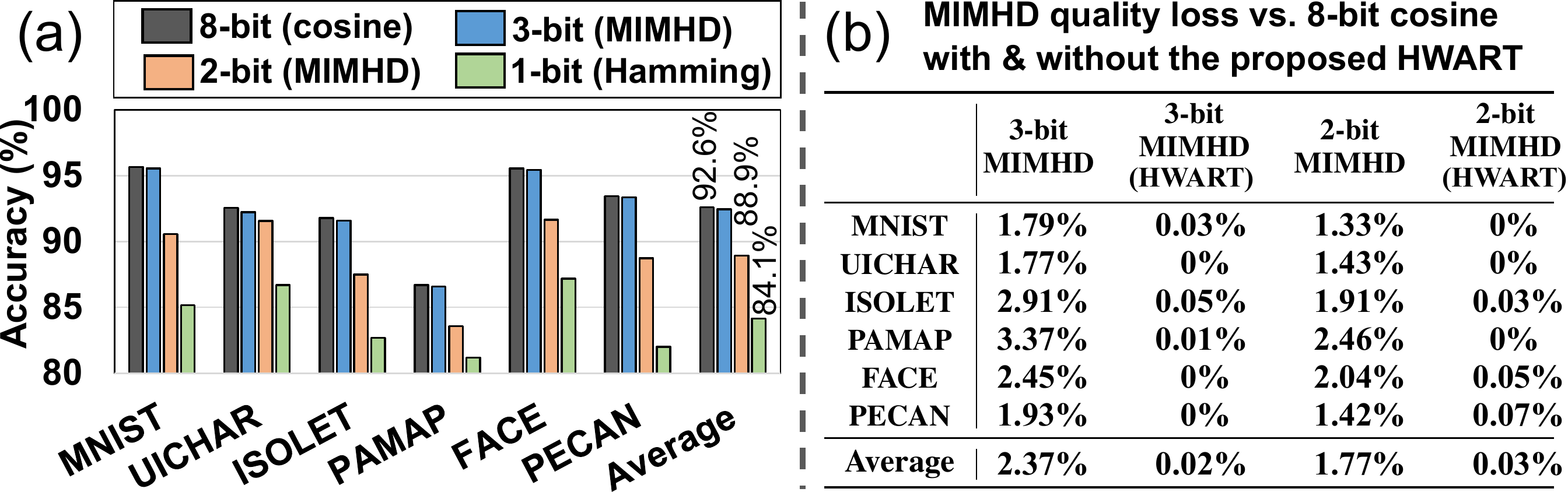, width=1\columnwidth}
\vspace{-6mm}
\caption{(a) Impact of HV precision on HDC classification accuracy, (b) \Design quality loss vs. cosine metric with and without HWART retraining.}
\vspace{-4mm}
\label{fig:accuracy}
\end{figure}

Fig.~\ref{fig:accuracy}b shows the quality loss of models trained with cosine distance and deployed on \Design, due to \Design's using MCAM distance metric. The results indicate that, in the same bit precision without HWART, 2 and 3-bit \Design provides, on average, 1.77\% and 2.7\% lower accuracy than the 8-bit HDC model with cosine distance. Note that the accuracy of 2-bit and 3-bit \Design models are still 6.5\% and 3.2\% higher than the binary model using the Hamming distance metric. As Fig.~\ref{fig:accuracy}b shows, enhancing the model with HWART can compensate for the quality loss caused by the MCAM distance metric. For example, \Design 3-bit achieves a similar accuracy as HDC with 3 and 8-bit precisions using cosine distance. 

\begin{figure}[t!]
\centering
\epsfig{file=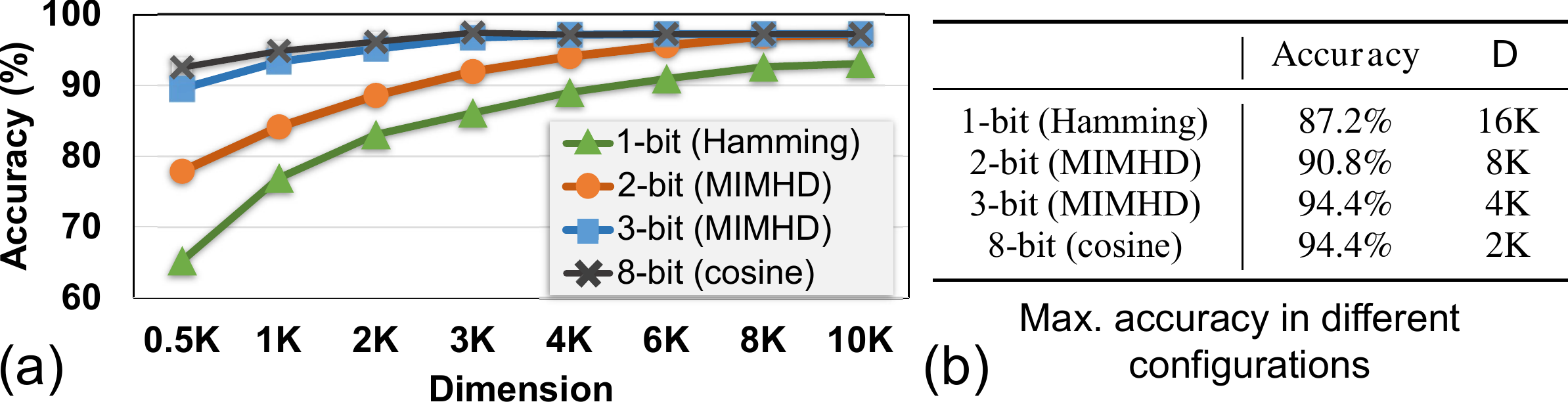, width=1\columnwidth}
\vspace{-6mm}
\caption{Impact of dimensionality on \Design accuracy.}
\vspace{-4mm}
\label{fig:dimension}
\end{figure} 

\subsection{\Design Efficiency}
We compare the performance of \Design with a NVIDIA GTX 1080 GPU in terms of energy and latency since GPUs are the most common computing fabrics for HDC. We compare \Design, which is entirely FeFET-based, in 1, 2, and 3-bit precision with the GPU to illustrate the efficiency trends of operating in different bit precisions with the same technology. We further compare the performance of \Design with state-of-the-art PIM-based HDC platforms: AHAM~\cite{imani2017exploring} that uses a Loser-Take All circuit to support Hamming distance search, and SearcHD~\cite{imani2019searchd} that exploits ganged-inverters and inverse CAM to support the same search functionality. Our evaluations are performed in two setups: (i) Maximum Accuracy, at maximum achievable accuracy of the designs, and (ii) Same Accuracy, when all designs achieve the same accuracy, i.e., the maximum accuracy of a binary design.

\subsubsection{\textbf{Energy}}
\begin{figure*}[t!]
\centering
\epsfig{file=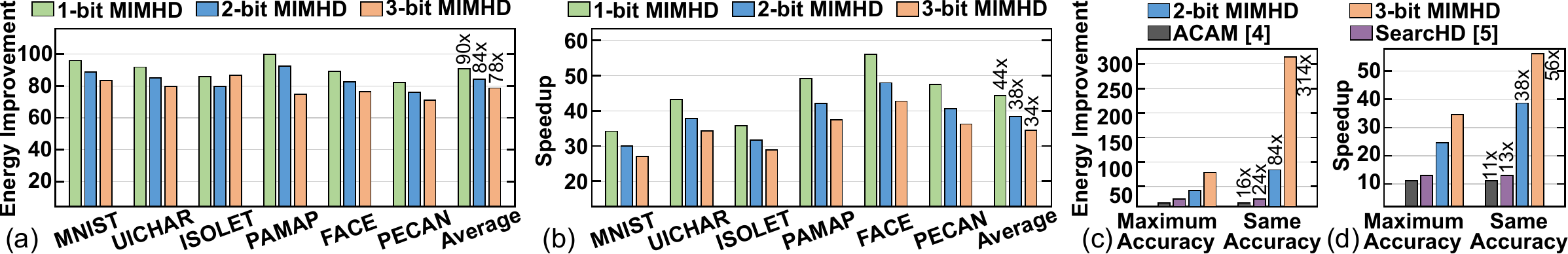, width=1\textwidth}
\vspace{-6mm}
\caption{(a) and (b) show the energy improvement and speedup of \Design over the GPU for $\mathcal{D}=4000$; (c) and (d) compare \Design with state-of-the-art binary HDC accelerators in maximum accuracy and same accuracy setups in terms of energy improvement and speedup over the GPU.}
\vspace{-5mm}
\label{fig:efficiency}
\end{figure*}

Fig.~\ref{fig:efficiency}a shows the energy improvement results of \Design over the GPU in different bit precisions and for different tasks (Table~\ref{tab:benchmark}). Due to its memory-centric nature, \Design in all bit precisions, achieves high energy improvements over the GPU by performing computations in-memory. The difference in energy improvement of \Design in different bit precisions is mainly due to the overhead of the ADCs as higher precision ADCs consume higher energy. On average, for the considered tasks, \Design using 1, 2, and 3 bits achieves $90\times$, $84\times$, and $78\times$ improvements over the GPU.

Fig.~\ref{fig:efficiency}c compares the average energy improvement of AHAM~\cite{imani2017exploring}, SearcHD~\cite{imani2019searchd}, and 2 and 3-bit \Design over the GPU. The most important factor that affects these results is the dimensionality. At Maximum Accuracy, 3-bit \Design operates in $\mathcal{D}{=}4k$ while AHAM and SearcHD operate in $\mathcal{D}{=}16k$ (Fig.~\ref{fig:dimension}b), but still \Design achieves a higher accuracy. Dimensionality has a linear relation with energy efficiency as it requires more resources, e.g., crossbar arrays. At Same Accuracy, the 3-bit \Design outperforms other implementations by a large margin as it only needs $\mathcal{D}{=}1k$ to achieve the same accuracy as binary implementations. The 3-bit \Design outperforms the state of the art, i.e., SearcHD, by $3.27\times$ and $13\times$ in Max Accuracy and Same Accuracy setups, respectively.

\subsubsection{\textbf{Latency}}
Fig.~\ref{fig:efficiency}b shows the speedup results of \Design in different bit precisions over the GPU for the tasks in Table~\ref{tab:benchmark}. \Design leverages highly parallel computation for encoding and associative search to achieve high speedups over the GPU. The encoding is performed in $n$ crossbar arrays in parallel (Fig.~\ref{fig:architecture}a). The single-step MCAM associative search increases speedup by alleviating memory access overheads and searching across all classes in one pass. The difference in speedup of \Design in different bit precisions is mainly due to the overhead of the ADCs. For the considered tasks, \Design with 1, 2, and 3-bit precisions achieves average improvements of $44\times$, $38\times$, and $34\times$ over the GPU, respectively.

Fig.~\ref{fig:efficiency}d compares the average speedup of AHAM~\cite{imani2017exploring}, SearcHD~\cite{imani2019searchd}, and \Design with 2 and 3-bit precisions over the GPU. Unlike energy improvement, speedup does not scale linearly with respect to dimensionality, since a significant portion of the computation happens in parallel. However, lower dimensinality reduces some of the peripheral overhead which is reflected in the results. 3-bit \Design outperforms the state of the art, i.e., SearcHD, by $2.6\times$ and $4.3\times$ in Maximum Accuracy and Same Accuracy setups, respectively.


\subsection{\Design Robustness}
The technological and fabrication issues in highly scaled technology nodes add a significant amount of noise to both memory and computing units~\cite{rahmani2016reliability, ISSCC}. These issues add extra computational error sources, which degrades the quality of inference. One of the main advantages of HDC is its high robustness to noise and failure. In HDC, HVs are random and holographic with i.i.d. components. Each HV stores the information across all its components so that no component is more responsible for storing any piece of information than another. This makes HDC robust against computational errors. However, HDC with floating point precision HVs has a much lower robustness to hardware noise compared to HDC with binary precision HVs. In high precision representation, a bit-flip due to noise in the exponent, the most significant bits of mantissa, or the sign bit can drastically change weight values, thus changing the prediction results. 

Models deployed on \Design inherit the intrinsic robustness of HDC. Table~\ref{tab:robustness} reports the quality loss of HDC using different bit precisions ($D{=}4k$). Our evaluation indicates that HDC quality loss increases with HV precision. For example, using a 10\% random bit flip (noise) in the encoding and associative search modules, 2 and 8-bit HDC experience 4.1\% and 12.4\% quality loss, respectively. Thus, it is desired to operate in lower precisions. \Design enables highly robust HDC inference by operating in low-precisions (1, 2, and 3 bits). HDC robustness also depends on HV dimensionality. Increasing dimensionality from $D{=}4k$ to $D{=}10k$ leads to 3.4$\times$ lower noise sensitivity. However, this boost in the robustness comes at the expense of lower computation efficiency. Note that HDC is significantly less sensitive than the existing learning solutions operating in floating point precision. For example, considering 2\% and 10\% random noise in model parameters, neural networks exhibit 9.4\% and 40.0\% quality loss, respectively~\cite{cano2021onlinehd}.

\begin{table}
\caption{HDC quality loss using different HV precisions.}
\vspace{-2mm}
\centering
\resizebox{0.9\columnwidth}{!}{
\begin{tabular}{cc|ccccc}
\toprule
\multicolumn{2}{c|}{\textit{\textbf{Hardware Error}}}                     & \textit{\textbf{1\%}} & \textit{\textbf{2\%}} & \textit{\textbf{5\%}} & \textit{\textbf{10\%}} & \textit{\textbf{15\%}} \\ \midrule 
\multicolumn{1}{c}{\multirow{4}{*}{\textbf{\Design}}} & \textbf{1-bit}   & 0.0\%                 & 0.0\%                 & 0.9\%                 & 3.1\%                  & 5.2\%                  \\ 
\multicolumn{1}{c}{}                                 & \textbf{2-bit}  & 0.0\%                 & 0.3\%                 & 1.2\%                 & 4.1\%                  & 6.7\%                  \\ 
\multicolumn{1}{c}{}                                 & \textbf{3-bit}  & 0.1\%                 & 0.3\%                 & 2.0\%                 & 5.9\%                  & 9.3\%                 \\ 
\multicolumn{1}{c}{}                                 & \textbf{8-bit}  & 1.2\%                 & 3.7\%                 & 5.5\%                 & 12.4\%                 & 18.7\%                 \\ \bottomrule
\end{tabular}
}\label{tab:robustness}
\vspace{-5mm}
\end{table}

\section{Conclusion}
We proposed \Design, accurate and efficient in-memory implementation of HDC inference. \Design uses multi-bit values to represent HV dimensions and exploits FeFET crossbar arrays and FeFET MCAM arrays to perform in-memory encoding and associative search. We also proposed a novel HWART algorithm which iteratively retrains the HDC model to compensate for possible accuracy loss due to the MCAM distance metric. Our evaluation shows that \Design provides comparable accuracy to high precision models while enabling higher computation efficiency.

\section*{Acknowledgment}
This work was supported in part by Semiconductor Research Corporation (SRC) Task No. 2988.001, ASCENT, Department of the Navy, Office of Naval Research, grant \#N00014-21-1-2225, and a generous gift from Cisco.

{\scriptsize
\bibliographystyle{./my_bib_style.bst}
\bibliography{mybibliography}}

\begin{thebibliography}{10}

\bibitem{kanerva2009hyperdimensional}
P.~Kanerva.
\newblock Hyperdimensional computing.
\newblock {\em Cognn Comput.}, 2009.

\bibitem{mitrokhin2019learning}
A.~Mitrokhin et~al.
\newblock Learning sensorimotor control with neuromorphic sensors.
\newblock {\em Science Robotics}, 4(30), 2019.

\bibitem{kleyko2018classification}
D.~Kleyko, et~al.
\newblock Classification and recall with binary hyperdimensional computing.
\newblock {\em IEEE Trans Neural Netw Learn Syst}, (99):1--19, 2018.

\bibitem{Rasanen15}
O.~Rasanen et~al.
\newblock Sequence prediction with sparse distributed hyperdimensional coding
  applied to the analysis of mobile phone use patterns.
\newblock {\em IEEE Trans. Neural Netw. Learn. Syst.}, PP(99):1--12, 2015.

\bibitem{rahimi2018efficient}
A.~Rahimi, et~al.
\newblock Efficient biosignal processing using hyperdimensional computing.
\newblock {\em Proceedings of the IEEE}, 107(1):123--143, 2018.

\bibitem{imani2019framework}
M.~Imani et~al.
\newblock A framework for collaborative learning in secure high-dimensional
  space.
\newblock In {\em CLOUD}, pages 435--446. IEEE, 2019.

\bibitem{imani2021revisiting}
M.~Imani et~al.
\newblock Revisiting hyperdimensional learning for fpga and low-power
  architectures.
\newblock In {\em HPCA}, 2021.

\bibitem{li2016hyperdimensional}
H.~Li et~al.
\newblock Hyperdimensional computing with 3d vrram in-memory kernels.
\newblock In {\em IEDM}. IEEE, 2016.

\bibitem{imani2017exploring}
M.~Imani et~al.
\newblock Exploring hyperdimensional associative memory.
\newblock In {\em HPCA}, pages 445--456. IEEE, 2017.

\bibitem{imani2019quanthd}
M.~Imani et~al.
\newblock Quanthd: A quantization framework for hyperdimensional computing.
\newblock {\em TCAD}, 2019.

\bibitem{cano2021onlinehd}
A.~Hernandez-Cano et~al.
\newblock Onlinehd: Robust, efficient, and single-pass online learning using
  hyperdimensional system.
\newblock In {\em DATE}, 2021.

\bibitem{bartol2015nanoconnectomic}
T.~M. Bartol~Jr, et~al.
\newblock Nanoconnectomic upper bound on the variability of synaptic
  plasticity.
\newblock {\em Elife}, 4:e10778, 2015.

\bibitem{kazemi2021memory}
A.~Kazemi, et~al.
\newblock In-memory nearest neighbor search with fefet multi-bit
  content-addressable memories.
\newblock In {\em DATE}. IEEE, 2021.

\bibitem{jerry2018ferroelectric}
M.~Jerry, et~al.
\newblock A ferroelectric field effect transistor based synaptic weight cell.
\newblock {\em Journal of Physics D: Applied Physics}, 2018.

\bibitem{imani2019searchd}
M.~Imani et~al.
\newblock Searchd: A memory-centric hyperdimensional computing with stochastic
  training.
\newblock {\em TCAD}, 2019.

\bibitem{nazemi2020synergiclearning}
M.~Nazemi, et~al.
\newblock Synergiclearning: Neural network-based feature extraction for
  highly-accurate hyperdimensional learning.
\newblock {\em arXiv}, 2020.

\bibitem{Poduval2021cognitive}
P.~Poduval et~al.
\newblock Cognitive correlative encoding for genome sequence matching in
  hyperdimensional system.
\newblock In {\em DAC}, 2021.

\bibitem{Cano2021reghd}
A.~Hérnandez-Cano et~al.
\newblock Reghd: Robust and efficient regression in hyper-dimensional learning
  system.
\newblock In {\em DAC}, 2021.

\bibitem{karunaratne2020memory}
G.~Karunaratne et~al.
\newblock In-memory hyperdimensional computing.
\newblock {\em Nature Electronics}, pages 1--11, 2020.

\bibitem{ni2018circuit}
K.~Ni, et~al.
\newblock A circuit compatible accurate compact model for ferroelectric-fets.
\newblock In {\em IEEE Symposium on VLSI Technology}, 2018.

\bibitem{kazemi2020hybrid}
A.~{Kazemi}, et~al.
\newblock A hybrid femfet-cmos analog synapse circuit for neural network
  training and inference.
\newblock In {\em ISCAS}, pages 1--5, 2020.

\bibitem{ni2019ferroelectric}
K.~Ni et~al.
\newblock Ferroelectric ternary content-addressable memory for one-shot
  learning.
\newblock {\em Nature Electronics}, 2(11):521--529, 2019.

\bibitem{abadi2016tensorflow}
M.~Abadi et~al.
\newblock Tensorflow: Large-scale machine learning on heterogeneous distributed
  systems.
\newblock {\em arXiv preprint arXiv:1603.04467}, 2016.

\bibitem{peng2019dnn+}
X.~Peng, et~al.
\newblock Dnn+ neurosim: An end-to-end benchmarking framework for
  compute-in-memory accelerators with versatile device technologies.
\newblock In {\em IEDM}, pages 32--5. IEEE, 2019.

\bibitem{lecun1998gradient}
Y.~LeCun et~al.
\newblock Gradient-based learning applied to document recognition.
\newblock {\em Proceedings of the IEEE}, 86(11):2278--2324, 1998.

\bibitem{ciregan2012multi}
D.~Ciregan et~al.
\newblock Multi-column deep neural networks for image classification.
\newblock In {\em CVPR}, pages 3642--3649. IEEE, 2012.

\bibitem{anguita2012human}
D.~Anguita et~al.
\newblock Human activity recognition on smartphones using a multiclass
  hardware-friendly support vector machine.
\newblock In {\em AAL}, pages 216--223. Springer, 2012.

\bibitem{Dua:2019}
D.~Dua et~al.
\newblock {UCI} machine learning repository, 2017.

\bibitem{reiss2012introducing}
A.~Reiss et~al.
\newblock Introducing a new benchmarked dataset for activity monitoring.
\newblock In {\em ISWC}, pages 108--109. IEEE, 2012.

\bibitem{angelova2005pruning}
A.~Angelova et~al.
\newblock Pruning training sets for learning of object categories.
\newblock In {\em CVPR}. IEEE, 2005.

\bibitem{shome}
Pecan street dataport.
\newblock \url{https://dataport.cloud/}.

\bibitem{rahmani2016reliability}
A.~Rahmani et~al.
\newblock Reliability-aware runtime power management for many-core systems in
  the dark silicon era.
\newblock {\em TVLSI}, 25(2):427--440, 2016.

\bibitem{ISSCC}
T.~Wu et~al.
\newblock Brain-inspired computing exploiting carbon nanotube fets and
  resistive ram.
\newblock In {\em ISSCC}. IEEE, 2018.

\end{thebibliography}

\end{document}